\documentclass[12pt]{article}
\usepackage[english]{babel}
\usepackage{amsfonts}
\usepackage{amsmath}
\usepackage{amssymb}
\usepackage{mathrsfs}
\usepackage{multicol}
\usepackage{fancybox}
\usepackage{dsfont}
\usepackage{color}
\usepackage{hyperref}
\usepackage{subfigure} 
\usepackage{graphicx} 
\usepackage{caption}
\usepackage{lineno}

\def\be{\begin{equation}}
\def\ee{\end{equation}}
\def\tl{\tilde} 
 
\def\gm{\gamma} 
\def\lm{\lambda}

\def\tr{\textrm{tr}}

\def\d'{``}

\newtheorem{th1}{Theorem}

\def\be{\begin{equation}}
\def\ee{\end{equation}}
\def\bea{\begin{eqnarray}}
\def\eea{\end{eqnarray}}

\newcommand{\Da}{{\mathcal D}}
\newcommand{\da}{{\mathfrak d}}
\def\dblone{\hbox{$1\hskip -1.2pt\vrule depth 0pt height 1.6ex width 0.7pt \vrule depth 0pt height 0.3pt width 0.12em$}}
\def\d'{``}

\begin{document}

\begin{center}
\Large{\bf{B\"acklund transformations and Hamiltonian flows}}
\end{center}

\begin{center}
{ {  Federico Zullo}}

{ School of Mathematics, Statistics and Actuarial Science,\\ University of Kent, CT2 7NF, Canterbury, Kent, UK.
\\~~E-mail: F.Zullo@kent.ac.uk}

\end{center}

\medskip
\medskip

\begin{abstract}
\noindent
In this work we show that, under certain conditions, parametric B\"acklund transformations (BTs) for a finite dimensional integrable system can be interpreted as solutions to the equations of motion defined by an associated non-autonomous Hamiltonian. The two systems share the same constants of motion. This observation leads to the identification of the Hamiltonian interpolating the iteration of the discrete map defined by the transformations, that indeed in numerical applications can be considered a linear combination of the integrals appearing in the spectral curve of the Lax matrix. An example with the periodic Toda lattice is given. 
\end{abstract}

\bigskip\bigskip

\noindent


\section{Introduction} \label{intr}
In recent years there has been a huge interest in the theory of integrable discretization. Given a Poisson manifold $\mathcal{P}$ with a Poisson bracket $\{\cdot, \cdot\}$ and a Hamiltonian function $H$ of a Liouville integrable system, the flow on $\mathcal{P}$ is given by:
\be\label{intdis}
\dot{x}_{i}=f_i(x)=\{H,x_i\}.
\ee

An \d'integrable discretization'' usually means (see e.g. \cite{Sbook}) a discrete analogue of the Hamiltonian flow (\ref{intdis}), that is a map $\mathcal{P}\to \mathcal{P}$,
$$
\tl{x}_i=\omega_i(\tl{x},x,\mu),
$$
depending on some parameter $\mu$ and possessing the following characteristics:
\begin{itemize}
\item It approximates, at least to first order in $\mu$, the flow (\ref{intdis}):
$$
\tl{x}_i=x_i+\mu f_i(x)+O(\mu^2).
$$
\item It preserves the Poisson structure in the limit $\mu\to 0$:
$$
\{\tl{x}_i,\tl{x}_j\}=\widetilde{\{x_i,x_j\}}+O(\mu).
$$
\item It is integrable, that is it possesses a sufficient number of conserved quantities independent and in involution, approximating the integrals $H_k$ of the continuous model to first order in $\mu$:
$$
\tl{H}_k=H_k+O(\mu).
$$
\end{itemize}

A significant number of results have been published
on the discretization of finite dimensional systems fitting the previous definition \cite{BLS} \cite{BS} \cite{BSbook} \cite{HP} \cite{MQ} \cite{MV} \cite{NRK} \cite{PPS} \cite{RZT} \cite{Z}  (see also \cite{Sbook} and references therein). Among the various techniques, the approach using the B\"acklund transformations (BTs) appears noteworthy: indeed, despite the fact that it is non-algorithmic (unlike other methods, e.g. the splitting method \cite{MQ}), it possesses characteristics, such as explicitness, exact preservation of all the integrals and canonicity (see e.g. \cite{HKR}), that deserve special consideration. 

The aim of this paper is to show how, provided the \d'spectrality property'' \cite{KS} holds, BTs for finite dimensional systems represent an \emph{exact} discretization of the underlying continuous integrable model. With the adjective \emph{exact}  we mean the equivalence, at all times and at all orders, of the trajectories of the discrete flow defined by the transformations and of the trajectories of the continuous flow of the model.

In Section \ref{sec1} we will consider an integrable system described by a Lax matrix possessing a parameter family of BTs with the spectrality property. We will show how these transformations can be seen as the integral curves of a system of non-autonomous Hamiltonian equations sharing the same conserved quantities with the original model. Then it will be pointed out that indeed the recurrences defined by the transformations lie on the trajectories of the continuous system defined by the Hamiltonian
\be
\mathcal{H}=\int_{0}^\mu \Phi(H_k,\mu,\{\lm_i\}) d\mu ,
\ee
where $\Phi$ is a function related to the spectrality property, $\mu$ is a parameter of the transformations playing the role of a time, $H_k$ are the conserved quantities of the continuous model and $\{\lm_i\}$ represents a set of parameters on which the BTs depend.  At the end of the Section we will discuss the usefulness of the formulae obtained. In Section \ref{sec2} we will give an application of what stated in Section \ref{sec1} with a two-parameters BTs on the periodic Toda lattice. The last Subsection will present some numerical examples of this model.

\section{BTs \& integrable discretization: a Hamiltonian approach}\label{sec1}
Consider an integrable system with a Lax matrix $L(\lm)$, where $\lm$ is the spectral parameter. 
The conserved quantities of the model, independent and in involution, can be obtained as the coefficients of an expansion in $\lm$ of the eigenvalues of $L(\lm)$ (see e.g. \cite{BBT}). So the characteristic equation
\begin{equation}\label{speccurve}
\det(L(\lm)-v\dblone)=0, 
\end{equation}
defines the eigenvalues of $L(\lm)$ in terms of the conserved quantities and $\lm$: this corresponds to the well known isospectral evolution of the eigenvalues of a Lax matrix \cite{BBT}. The curve defined by $\det(L(\lm)-v\dblone)=0$ is time-independent: it can be seen as a Riemann surface and it is called the \emph{spectral curve}.

It is convenient to think about our model as written in terms of canonical variables, say $\{p_i, q_i\}_{i=1}^N$, so that we have the usual Poisson brackets, that is
$$
\{p_i,q_j\}=\delta_{ij},\qquad \{p_i,p_j\}=\{q_i,q_j\}=0,
$$
and the involutivity conditions on the conserved quantities $H_i$, i.e.
\begin{equation}\label{Hj}
\{H_i,H_j\}=0, \qquad \forall i,j=1,...,N.
\end{equation}

A set of BTs for the system is simply a set of canonical transformations 
\begin{equation}\label{backlund}(p_i,q_i)\overset{BT}{\Longrightarrow}(\tl{p}_i,\tl{q}_i), \qquad \textrm{where}\qquad \left\{\begin{aligned}
&\tl{p}_i=f_i(p,q) \\
&\tl{q}_i=g_i(p,q) 
\end{aligned}\right.
\end{equation}
from $(p_i, q_i)_{i=1}^N$ to a new set $(\tl{p}_i, \tl{q}_i)_{i=1}^N$ such that the $N$ functions $(H_i)_{i=1}^N$ are invariant under the transformations, that is $H_i(\tl{p},\tl{q})=H_i(p,q)$. To simplify the notation we will drop hereafter the subscripts in the independent variables as in the formula above. From this definition the classical property of BTs (or better auto-BTs) that they send solutions of the equations of motion into solutions is easily proved, since the variables with the tilde and those without the tilde satisfy the same system of differential equations. The application of BTs in numerical analysis comes from the observation that actually the relations (\ref{backlund}) define explicit recurrences just by thinking of variables with tilde as the same as those without tilde but computed at the next time step \cite{LB}:
\begin{equation}\label{rec}\begin{aligned}
&(p_i)_{n+1}=f_i((p)_n,(q)_n), \\
&(q_i)_{n+1}=g_i((p)_n,(q)_n).
\end{aligned}
\end{equation}
In the previous notation $p_i$ is equivalent to $(p_i)_0$, $\tl{p}_i$ is equivalent to $(p_i)_1$ and so on.

As pointed out in \cite{KS}, the invariance of the conserved quantities, and hence of the spectrum of $L(\lm)$, implies the existence of a \emph{dressing} or \emph{Darboux} matrix $D(\lm)$ intertwining the two matrices, $L(\lm)$ and $\tl{L}(\lm)\doteq L(\lm,\tl{p},\tl{q})$, that is
\begin{equation}\label{intertwin}
\tl{L}(\lm)D(\lm)=D(\lm)L(\lm).
\end{equation}
Also, because the transformations are canonical, there exists a generating function, say $F(q,\tl{q})$, such that 
\begin{equation*} \begin{aligned}
&p_i=\frac{\partial F(q,\tl{q})}{\partial q_i},\\
&\tl{p}_i=-\frac{\partial F(q,\tl{q})}{\partial \tl{q}_i}.
\end{aligned}
\end{equation*} 

Of particular interest are the BTs depending on a parameter, say $\mu$. It is important to note that in this case it is always possible to connect the transformations to the identity, in the sense that for a particular value of the parameter, that can always be chosen to be the value zero, one has $\tl{p}_i=p_i$ and $\tl{q}_i=q_i$. Indeed, if the original transformations are not connected to the identity we can obtain a new set of transformations that, by construction, are the identity transformations for $\mu=0$, by the following argument (see also \cite{HKR}).

Let the dressing matrix defining the transformations be $D(\lm, \lm_1)$, where $\lm_1$ is the parameter of the transformations. Since BTs are canonical transformations we can compose a transformation with parameter $\lm_1$ and a transformation with parameter $\lm_2$ to obtain a two-parameter set of BTs. So composing the transformation with $\lm_1\doteq \lm_0+\mu $ with its inverse transformation but calculated at $\lm_2\doteq \lm_0-\mu $ will give a transformation that in the limit $\mu \to 0$ goes to the identity. The only question is whether the map can be inverted; but given $D(\lm,\lm_1)$, the inverse transformation is just found by considering the adjugate (the transpose of the cofactor matrix) of the matrix $D(\lm,\lm_1)$, say $\hat{D}(\lm,\lm_1)$, in equation (\ref{intertwin}). The dressing matrix for the composed transformation then will be $D(\lm,\lm_0,\mu)=\hat{D}(\lm,\lm_2)\cdot D(\lm,\lm_1)$. We will give an example of such a construction in Section \ref{sec2}.

It is quite natural to interpret the parameter $\mu$ as a sort of \d'time'' with respect to which the BTs evolve. As a matter of fact BTs are parametric canonical transformations; from this it follows that they are the integral curves (whit parameter $\mu$) of a Hamiltonian vector field since, at least locally, there exists a Hamiltonian $K(\tl{p},\tl{q},\mu)$ such that: 
\begin{equation}\label{K} \begin{aligned}
&\frac{\partial \tl{q}_i}{\partial \mu}=\frac{\partial K}{\partial \tl{p}_i},\\
&\frac{\partial \tl{p}_i}{\partial \mu}=-\frac{\partial K}{\partial \tl{q}_i}.
\end{aligned}
\end{equation} 
(See for example \cite{FM} for the proof). As we will now show, if the so-called \emph{spectrality property} \cite{KS} holds, it is possible to find the function $K(\tl{p},\tl{q},\mu)$ explicitly. This function is related to the eigenvalues of the Lax matrix and so contains all the conserved quantities of the system; also, it depends on the parameter $\mu$ of the transformations. BTs are then the integral curves of a non-autonomous Hamiltonian vector field, the time being $\mu$. In order to clarify the previous statement let us recall the definition of the spectrality property. Suppose we have a set of BTs depending on a parameter $\mu$:
\begin{equation*} \begin{aligned}
&\tl{p}_i=f_i(p,q,\mu),\\
&\tl{q}_i=g_i(p,q,\mu).
\end{aligned}
\end{equation*} 
By applying the inverse function theorem, there exists a generating function $F$ such that \cite{FM}
\begin{equation*} \begin{aligned}
&p_i=\frac{\partial F(q,\tl{q},\mu)}{\partial q_i},\\
&\tl{p}_i=-\frac{\partial F(q,\tl{q},\mu)}{\partial \tl{q}_i}.
\end{aligned}
\end{equation*}
Consider the canonically conjugate variable $\Phi$ with respect to $\mu$:
\begin{equation}\label{ccvtm}
\Phi \doteq\left.\frac{\partial F}{\partial \mu}\right|_{\tl{q}=\tl{q}(p,q,\mu)}.
\end{equation}
The spectrality property \cite{KS} says that for some function $\digamma$, the pair $(\mu,\digamma(\Phi))$ lies on the spectral curve (\ref{speccurve}): 
\begin{equation}\label{spe}
\Phi =\left.\frac{\partial F}{\partial \mu}\right|_{\tl{q}=\tl{q}(p,q,\mu)}\qquad \Longrightarrow\qquad \det(L(\mu)-\digamma(\Phi)\dblone)=0.
\end{equation}

As an example, we will see in Section \ref{sec2} that in the case of the Toda lattice the function $\digamma$ is just the logarithm. It seems that the spectrality property is \d'universal'' in the sense of being shared by a large class of models (for example Toda \cite{S}, Ruijsenaars \cite{NRK}, Henon-Heiles \cite{HKR1}, Gaudin \cite{HKR}, Kirchhoff top \cite{RZK}, Mumford systems \cite{KV}). 

According to some authors\footnote{See e.g. \cite{S} where the author writes \d'\emph{the property...needs more
research to uncover its algebraic and geometric meaning}''.} the spectrality property could be better understood by taking an algebraic, or geometric, point of view. Here we want just to give a Hamiltonian interpretation of this property. Our approach is based on two properties: 
\begin{enumerate} 
\item $\Phi$ is a function of the phase space variables but only as a combination of the conserved quantities, because it lies on the spectral curve.
\item The function $\Phi$ and the parameter $\mu$ are canonically conjugated, so it is possible to enlarge the phase space by adding the pair $(\Phi,\mu)$ to the state variables \cite{FM}; then we can write the form $dF$ as the difference between two Poincar\'e Cartan forms as follows: 
\begin{equation}\label{PoincareCartan}
dF=\sum_i p_idq_i-\tl{p}_id\tl{q}_i+\Phi d\mu.
\end{equation}
\end{enumerate}

Let us see the consequences of the previous two properties. Property 1 implies that the function $\Phi$ is invariant under the transformations, that is:
$$
\hat{\Phi}(\tl{p},\tl{q},\mu)=\Phi(p,q,\mu),\qquad  \textrm{where}\qquad \left\{\begin{aligned}
&\Phi(p,q,\mu) =\left.\frac{\partial F}{\partial \mu}\right|_{\tl{q}=\tl{q}(p,q,\mu)},\\
&\hat{\Phi}(\tl{p},\tl{q},\mu)=\left.\frac{\partial F}{\partial \mu}\right|_{q=q(\tl{p},\tl{q},\mu)}.
\end{aligned}\right.
$$
Property 2 implies that if $(p(\mu),q(\mu))$ are the integral curves generated by \emph{any} Hamiltonian $T(p,q,\mu)$, then $\tl{p}(\mu)$ and $\tl{q}(\mu)$ are the integral curves generated by the new Hamiltonian $K(\tl{p},\tl{q},\mu)$ given by: 
\begin{equation}\label{hamil}
K(\tl{p},\tl{q},\mu)=\hat{T}(p(\tl{p},\tl{q},\mu),q(\tl{p},\tl{q},\mu),\mu)+\hat{\Phi}(\tl{p},\tl{q},\mu),
\end{equation}  
where $\hat{T}$ is the transform of $T(p,q,\mu)$ when expressed in terms of $\tl{p}$ and $\tl{q}$ (see also \cite{FM}, Remark 10.13).

Let us make an example to clarify this last property. Take the canonical transformation $(\tl{p},\tl{q})\to (p,q)$ with its inverse  $(p,q)\to(\tl{p},\tl{q})$, defined by:
\begin{equation*} \begin{aligned}
&\tl{p}=p+\mu q,\\
&\tl{q}=q+\mu p+\mu^2 q.
\end{aligned}
\end{equation*}
The generating function of these transformations and the corresponding derivative of the generating function with respect to $\mu$ are
$$
F=-\frac{(\tl{q}-q)^2+q^2\mu^2}{2\mu},\qquad \frac{\partial F}{\partial \mu}=\frac{(\tl{q}-q)^2-q^2\mu^2}{2\mu^2}.
$$
So in this example the function $\hat{\Phi}$ is given by
$$
\hat{\Phi}(\tl{p},\tl{q},\mu)=\frac{\tl{p}^2-(\tl{q}-\mu \tl{p})^2}{2}.
$$
We can conclude that if $(p(\mu),q(\mu))$ are the integral curves generated by the (arbitrary) Hamiltonian $T(p,q,\mu)$, then $\tl{p}(\mu)$ and $\tl{q}(\mu)$ are the integral curves generated by the new Hamiltonian $K(\tl{p},\tl{q},\mu)$ given by 
\begin{equation*}
K(\tl{p},\tl{q},\mu)=\hat{T}(p(\tl{p},\tl{q},\mu),q(\tl{p},\tl{q},\mu),\mu)+\frac{\tl{p}^2-(\tl{q}-\mu \tl{p})^2}{2}.
\end{equation*}  
More explicitly, after finding the inverse transformations  $(p,q)\to(\tl{p},\tl{q})$, we can write
\begin{equation*}
K(\tl{p},\tl{q},\mu)=T(\tl{p}-\mu\tl{q}+\mu^2\tl{p},\tl{q}-\mu\tl{p},\mu)+\frac{\tl{p}^2-(\tl{q}-\mu \tl{p})^2}{2}.
\end{equation*} 

Note that in this example $\hat{\Phi}\neq \Phi$ since we considered canonical transformations, not BTs: roughly speaking the BTs are canonical transformations with the further constraint that $\hat{\Phi}=\Phi$.

Now let us make this further observation: as explained before, it is always possible to have a set of BTs connected to the identity, so that
\begin{equation}\label{ICI}
\left.\tl{p}_i\right|_{\mu=0}=p_i, \qquad \qquad \left.\tl{q}_i\right|_{\mu=0}=q_i.
\end{equation}
Since the function $T(p,q,\mu)$ is arbitrary, we can choose it to be zero. With this choice $(p_i,q_i)_{i=1}^N$ are by definition just constants, that from (\ref{ICI}) are identified to be the initial conditions. From  (\ref{hamil}) and (\ref{ICI}) then readily follows the following
\begin{th1}\label{th1}
Assuming the spectrality property (\ref{spe}) holds, the BTs $(\tl{p}_i(\mu),\tl{q}_i(\mu))_{i=1}^N$ are the integral curves of the following system of (non-autonomous) Hamiltonian equations:
\begin{equation}\label{x} \begin{aligned}
&\frac{\partial \tl{q}_i}{\partial \mu}=\frac{\partial \Phi(\tl{p},\tl{q},\mu)}{\partial \tl{p}_i},\\
&\frac{\partial \tl{p}_i}{\partial \mu}=-\frac{\partial \Phi(\tl{p},\tl{q},\mu)}{\partial \tl{q}_i},
\end{aligned}
\end{equation} 
corresponding to the initial conditions (\ref{ICI}).
\end{th1}

From this point of view obtaining a set of explicit BTs for an integrable model with the spectrality property, means having the explicit general solution of a non-autonomous Hamiltonian system sharing the same conserved quantities with the original model. So we are considering two set of flows: the first set is given by all the flows defined by the integrable model, the second set is given by the flow defined by $\Phi$. In the following we aim to clarify the relations among these flows in order to point out a concrete application of Theorem \ref{th1}.

The key point is that, because of the spectrality property, the flow defined by (\ref{x}) possesses the same conserved quantities (\ref{Hj}) as the corresponding integrable model defined by the Lax matrix $L(\lm)$. Since the model is integrable (in the sense of Liouville), there exist action-angle variables $(I_k,\theta_k)_{k=1}^N$, such that the action variables $I_k$ are functions only of the conserved quantities and the pairs $(I_k,\theta_k)$ are canonically conjugate \cite{FM}: \begin{equation}\label{AA}
\{I_k,\theta_j\}=\delta_{kj}, \qquad \{I_k,I_j\}=0,\qquad \{\theta_k,\theta_j\}=0.
\end{equation}
Actually we do not need that all the variables $\theta_j$ are really \d'angles'', we could have also a splitting among bounded and unbounded variables (say $\theta_k\in \mathbb{T}^{m}, k=1...m$ and $\theta_k\in \mathbb{R}^{N-m}, k=m+1...N$). The key requirement is the canonicity (\ref{AA}) (together with the conservation of each of $I_j$). Having this in mind, for the sake of simplicity we continue to use indiscriminately the term angle in what follows.

Let us consider the flow (\ref{x}) defined by $\Phi$ in these coordinates. When expressed in terms of the set $(I_k,\theta_k)_{k=1}^N$ the Hamiltonian $\Phi(\tl{p},\tl{q},\mu)=\Phi(p,q,\mu)$ reduces to a function of action coordinates and $\mu$ only:
$$
\Phi(p,q,\mu)=\breve{\Phi}(I,\mu).   
$$ 

The system (\ref{x}) then becomes
\begin{equation} \label{vgf}\begin{aligned}
&\frac{\partial \tl{I}_j}{\partial \mu}=-\frac{\partial \breve{\Phi}(\tl{I},\mu)}{\partial \tl{\theta}_j}=0\Rightarrow \tl{I}_j=\textrm{const.}=I_{j},\\
&\frac{\partial \tl{\theta}_j}{\partial \mu}=\frac{\partial \breve{\Phi}(\tl{I},\mu)}{\partial \tl{I}_j}\Rightarrow \tl{\theta}_j=\theta_j+\frac{\partial}{\partial \tl{I}_j}\int_{0}^{\mu}\breve{\Phi}(\tl{I},\mu)d\mu ,
\end{aligned}
\end{equation} 
where $\tl{I}_j$ and $\tl{\theta}_j$ are the functions $I_j(p,q)$ and $\theta_j(p,q)$ evaluated in $p=\tl{p}$ and $q=\tl{q}$, i.e. $\tl{I}_j=I_j(\tl{p},\tl{q})$, $\tl{\theta}_j=\theta_j(\tl{p},\tl{q})$. In these coordinates the recurrences (\ref{rec}) are linearised: indeed from equations (\ref{vgf}) it follows that their \emph{explicit} solution is given by:
\begin{equation}\label{discr} \begin{aligned}
&(I_j)_n=(I_j)_0 ,\\
&(\theta_j)_{n}=(\theta_j)_0+n \frac{\partial}{\partial \tl{I}_j}\int_{0}^{\mu}\breve{\Phi}(\tl{I},\mu)d\mu .
\end{aligned}
\end{equation}
Replacing $n$ by $t$ we see that equations (\ref{discr}) represent the exact time discretization of the continuous system governed by the Hamiltonian:
\begin{equation}\label{Hdiscr}
\mathcal{H}=\int_{0}^{\mu}\breve{\Phi}d\mu .
\end{equation} 

Notice that there is no need to find the action-angle coordinates explicitly. The result (\ref{Hdiscr}) is independent of the choice of coordinates and can be restated as follows
\begin{th1}\label{th2} 
If the spectrality property holds, the discrete trajectories defined by the iteration of the recurrences
\begin{equation}\label{recus}\begin{aligned}
&(p_i)_{n+1}=f_i(p_n,q_n,\mu), \\
&(q_i)_{n+1}=g_i(p_n,q_n,\mu),
\end{aligned}
\end{equation} 
given by the BTs coincide with the trajectories of the Hamiltonian system defined by the Hamiltonian
\begin{equation*}
\mathcal{H}=\int_{0}^{\mu}\Phi\, d\mu ,
\end{equation*} 
where $\Phi$ is defined by the spectrality property (\ref{spe}).
\end{th1}
So, on one hand the BTs are the integral curves of the system of Hamiltonian equations generated by $\Phi$, but on the other hand they are the exact time discretization (i.e. preserving the trajectories) of the Hamiltonian equations generated by $\mathcal{H}=\int \Phi d\mu$. We stress again that 
$\Phi$ is a function of all the conserved quantities $H_i$ (\ref{Hj}) of the model.

The BTs, from a numerical point of view, are the discretization of the flows corresponding to \emph{linear} combinations of these flows generated by $H_i$ in the following sense: for any function $\mathfrak{F}(p,q)$ on the phase space we can write:  
\begin{equation}\label{linearcomb}
\frac{\partial \mathfrak{F}}{\partial t}=\{\mathcal{H},\mathfrak{F}\}=\sum_k c_k\{H_k,\mathfrak{F}\}, \quad \textrm{where}\quad c_k=\frac{\partial}{\partial H_k}\int_{0}^{\mu} \Phi d\mu .
\end{equation}
If we want to use the BTs concretely as a discretization tool for computer simulations, we need to fix the values of the initial conditions and the values of the parameters. Then we can start to iterate the map according to (\ref{recus}). Thanks to Theorem \ref{th2}, the result will be a sampling of the particular trajectory of the continuous flow (\ref{linearcomb}) corresponding to the same initial conditions. Note that the functions $c_k$ depend only on the conserved quantities and on the parameters of the transformations, so they are constants along the trajectories of the model, i.e. they are constants of motion.

Let us make some remarks about the applications of the previous formulae. Having in mind a physical model, usually one looks only at a discretization in some particular direction, corresponding to a given choice of the constants $c_k$ appearing in  (\ref{linearcomb}). So the problem is about the freedom we have to choose the values of these constants. 

The point is that the $c_k$ depend on the parameters of the transformations: if we have enough parameters, by picking carefully their values it is possible to obtain the particular combination of the constants $c_k$ we need. 
The problem is how to find the parameters $\lm_i$ as functions of given constants $c_k$:
\begin{equation}
c_k=\frac{\partial}{\partial H_k}\int_{0}^{\mu} \Phi(H_i,\mu,\{\lm_i\}) d\mu \quad \Longrightarrow \{\lm_i\}=\{\lm_i(c_k)\}.
\end{equation} 
Notice that \emph{for any fixed orbit} and for a given set of the constants $c_k$, the above integrals can be inverted numerically. An
example will be given in Subsection \ref{Numerics}. 

A final remark to make here is that the parameter $\mu$ plays the role of a time as in Theorem \ref{th1}, but also smaller values of the parameter $\mu$ correspond to smaller values of the time step.

\section{An example from the Toda lattice}\label{sec2}
In the rest of the paper we will present an application of the formulae discussed above to the Toda lattice. We summarize the main results of the Section:
\begin{enumerate}
\item A set of two-parameter BTs is found, as given in eqs. (\ref{D}) and (\ref{imptransf}) below. The maps are connected to the identity, are explicit, see eq. (\ref{xy}) below, and possess the spectrality property (see (\ref{newkmu})).
\item Applying Theorem \ref{th2} to the case $N=2$, it is shown how BTs correspond to addition formulae for special functions (elliptic functions in this case); see eqs. (\ref{linBts}) and (\ref{addform}). 
\item A concrete numerical application of Theorem \ref{th2} to the less trivial case $N=3$ is given in Subsection \ref{Numerics}, where particular flows, corresponding to fixed values of the constants $c_1$, $c_2$ and $c_3$ given by (\ref{eqdisc}), are discretised.     
\end{enumerate}

We assume periodic boundary conditions so that every lattice index $j$ can be substituted with the lattice index $j+N$. The Lax matrix of the model is given by a product of local Lax matrices $l_j(\lm)$ \cite{S}:
\begin{equation}\label{eq:lax} 
L(\lm)\doteq \left( \begin{array}{cc} A(\lm) & B(\lm)\\ C(\lm)& D(\lm)\end{array} 
\right)=l_{N}(\lm)\cdots l_{1}(\lm), \qquad l_{j}(\lm) \doteq  \left( \begin{array}{cc} \lm +p_j & -e^{q_j}\\ e^{-q_j}& 0\end{array} 
\right). 
\end{equation}
Recall that \d'local'' means that $l_j$ only depends on $p_j$ and $q_j$. The variables $(p_j,q_j)_{j=1}^{N}$ are pairs of canonically conjugate variables. 
The determinant of each local Lax matrix $l_j(\lm)$ is equal to 1, so the determinant of $L(\lm)$ is also equal to 1. The spectral curve in this case is a hyperelliptic curve: it is defined by the characteristic polynomial $\det (v-L(\lm))=0$ (see eq. (\ref{speccurve})), that is 
$$
v^2-\tr(L(\lm))v+1=0.
$$ 
This curve is constant with respect to the Toda flows: all the commuting Hamiltonians $H_j$ of the system can be obtained from the coefficients of the expansion of $\tr(L(\lm))$:
\be\label{tra}
\tr(L(\lm))=A(\lm)+D(\lm)=\sum_{k=0}^{N}H_k \lm^k.
\ee
Note that there are $N$ non-trivial Hamiltonians because $H_N=1$.

We now construct a class of BTs for this system following \cite{S}. 
Due to the splitting $L(\lm)=l_{N}(\lm)\cdots l_{1}(\lm)$ of the Lax matrix into local matrices, it is possible to recursively define a set of local matrices $d_j(\lm)$ by the relations
\be\label{local}
l_{j}(\lm,\tl{p}_j,\tl{q}_j)d_j(\lm)=d_{j+1}(\lm)l_j(\lm, p_j,q_j),
\ee    
so that the dressing matrix $D(\lm)$ (\ref{intertwin}) is given by $D(\lm)=d_1(\lm)$. 
The simplest choice for the matrices $d_j$ corresponds to a linear dependence on the spectral parameter \cite{S}:
\begin{equation}\label{dj} 
d_{j}(\lm) =  \left( \begin{array}{cc} \lm -\zeta -\alpha_j \beta_j & -\alpha_j \\ \beta_j & 1\end{array} 
\right), 
\end{equation}
where $\zeta$ is a parameter and $\alpha_j$ and $\beta_j$ depend on the dynamical variables. The dependences of $\alpha_j$ and $\beta_j$ are immediately fixed by the off-diagonal part of (\ref{local}), giving:
\begin{equation}\label{alphabeta} \begin{aligned}
&\beta_{j+1}=e^{-\tl{q}_j},\\
&\alpha_j=e^{q_j}.
\end{aligned}
\end{equation}
With these expressions, the similarity transformation (\ref{local}) defines the following \emph{implicit} set of BTs:
\begin{equation}\label{ptp} \begin{aligned}
&\tl{p}_{j}=-\left(e^{\tl{q}_j-q_j}+e^{q_{j+1}-\tl{q}_j}+\zeta\right),\\
&p_{j}=-\left(e^{\tl{q}_j-q_j}+e^{q_{j}-\tl{q}_{j-1}}+\zeta\right).
\end{aligned}
\end{equation}

However, the flow defined by (\ref{ptp}) is not connected to the identity. In order to obtain a flow connected to the identity, as explained in Section \ref{sec1}, we can compose two BTs, the first with parameter $\zeta=\lambda_1$ and the second given by the inverse transformations of the first one but evaluated at $\zeta=\lm_2$. Then when $\lm_2\to \lm_1$ the identity map is recovered. The details are given in Appendix \ref{Appendix}. From now on we consider directly the transformations given by the Darboux matrix (\ref{Dapp}) obtained in Appendix \ref{Appendix}, that is
\begin{equation}\label{D} 
\Da(\lm,\lm_1,\lm_2) =  \left( \begin{array}{cc} \lm-\lm_1-xy & -x \\ y(\lm_1-\lm_2+xy) & \lm -\lm_2 +x y\end{array} 
\right), 
\end{equation}
so that we look at the relation
\be\label{new}
\tl{L}(\lm)\Da(\lm,\lm_1,\lm_2)=\Da(\lm,\lm_1,\lm_2)L(\lm).
\ee
The matrix (\ref{D}) is degenerate for $\lm=\lm_1$ and $\lm=\lm_2$. The kernels of $\Da$ at $\lm=\lm_1, \lm_2$ are given by $|\Omega_1>=(1,-y)^{T}$, $|\Omega_2>=(x,\lm_2-\lm_1-xy)^{T}$ respectively. These are also eigenvectors of $L(\lm_1)$ and $L(\lm_2)$ with eigenvalues $v(\lm_1)$ and $v(\lm_2)^{-1}$, so that
\be\label{l1l2}
L(\lm_1)|\Omega_1>=v(\lm_1)|\Omega_1>, \qquad L(\lm_2)|\Omega_2>=v(\lm_2)^{-1}|\Omega_2>.
\ee
From Appendix \ref{Appendix}, we can explicitly write:
\begin{equation}\label{v1v2}\begin{split}
&v(\lm_1)=\frac{A(\lm_1)+D(\lm_1)-\gm(\lm_1)}{2}, \qquad  v(\lm_2)=\frac{A(\lm_2)+D(\lm_2)+\gm(\lm_2)}{2},\\
&\gm(\lm)^2\doteq (A(\lm)+D(\lm))^2-4.
\end{split}\end{equation}
From relations (\ref{l1l2}) the variables $y$ and $x$ are then easily found:
\begin{equation}\label{xy} \begin{aligned}
&y=\frac{A(\lm_1)-D(\lm_1)+\gm(\lm_1)}{2B(\lm_1)},\\
&x=\frac{2B(\lm_1)B(\lm_2)(\lm_2-\lm_1)}{B(\lm_2)(A(\lm_1)-D(\lm_1)+\gm(\lm_1))-B(\lm_1)(A(\lm_2)-D(\lm_2)-\gm(\lm_2))}.
\end{aligned}
\end{equation}
The relations above enable us to obtain explicit BTs. Indeed, thanks to (\ref{xy}), the Darboux matrix $\Da$ depends only on the variables without tilde, so the relation $\tl{L}=\Da L\Da^{-1}$ gives explicit transformations. Also, by setting
\be\label{lam0mu}
\lm_1=\lm_0+\mu, \qquad \lm_2=\lm_0-\mu,
\ee
we see that in the limit $\mu\to 0$ the identity transformation is recovered.

\subsection{Generating functions and Hamiltonian flows}
In this Subsection we will find the generating function for the composed transformations produced by the Darboux matrix (\ref{D}); then we will check the spectrality property. Finally, an application of Theorem \ref{th1} and Theorem \ref{th2} will be given.

The generating function $F$ solves the system:
\be\label{GF}
p_{j}=\frac{\partial F}{\partial q_j}, \qquad \tl{p}_{j}=-\frac{\partial F}{\partial \tl{q}_j}.
\ee
Again we look at the relation among the local Lax matrices, that is
\be\label{nlocal}
l_{j}(\lm,\tl{p},\tl{q})\da_j(\lm)=\da_{j+1}(\lm)l_j(\lm, p,q),
\ee 
where the matrices $\da_j$ are given by
\begin{equation}\label{daj} 
\da_j(\lm) =  \left( \begin{array}{cc} \lm-\lm_1 -x_j y_j & -x_j \\ y_j(\lm_1-\lm_2+x_jy_j) & \lm -\lm_2 +x_j y_j\end{array} 
\right). 
\end{equation}
The quantities $x_j$ and $y_j$ are to be determined. Recall that $\da_1=\Da$, as in (\ref{D}), so that $x_1=x$ and $y_1=y$. From the off-diagonal elements of (\ref{nlocal}) we get the expressions of $x_j$ and $y_j$ in terms of the dynamical variables $q_k$ and $\tl{q}_k$:
\be\label{xyqqt} 
x_j=e^{q_j}-e^{\tl{q}_j}, \qquad y_{j+1}(2\mu +x_{j+1}y_{j+1})=e^{-\tl{q}_j}-e^{-q_j}.
\ee
It is useful to introduce the variable $\eta_j$ defined by
$
\eta_j \doteq x_j y_j +\mu
$
. Indeed, from
$$
x_{j+1}y_{j+1}(2\mu +x_{j+1}y_{j+1})=\left(e^{q_{j+1}}-e^{\tl{q}_{j+1}}\right)\left(e^{-\tl{q}_{j}}-e^{-q_{j}}\right)\doteq w_{j+1},
$$
we see that $\eta_{j}^2=\mu^2+w_{j}$. With the help of these relations, the remaining equations (\ref{nlocal}) give
\begin{equation}\label{imptransf} \begin{aligned}
&\tl{p}_j=-\lm_0-\frac{e^{\tl{q}_j}\eta_j+e^{q_j}\eta_{j+1}}{e^{q_j}-e^{\tl{q}_j}},\\
&p_j=-\lm_0-\frac{e^{\tl{q}_j}\eta_{j+1}+e^{q_j}\eta_{j}}{e^{q_j}-e^{\tl{q}_j}}.
\end{aligned}
\end{equation}
After some calculations it is not difficult to check that the generating function $F$ for these transformations is given by
\be\label{newgf}
F=\sum_{k}\left(\mu \ln\left(\frac{\eta_k+\mu}{\eta_k-\mu}\right)-2\eta_k-\lm_0(q_k-\tl{q}_k)\right).
\ee

At this point we can calculate $\frac{\partial F}{\partial \mu}$ explicitly, as well as the function $\Phi$ in (\ref{ccvtm}). In the following we present the result immediately, providing the details in Appendix \ref{AppendixB}. 

By a direct calculation, the derivative of $F$ with respect to $\mu$ is
\be\label{vqqtmu}
\frac{\partial F}{\partial \mu}= \sum_k \ln\left(\frac{\eta_k+\mu}{\eta_k-\mu}\right).
\ee
The function $\Phi(p,q,\mu)$ is given by the following expression (see Appendix \ref{AppendixB}):
\be\label{newkmu}
\Phi=\left.\frac{\partial F}{\partial \mu}\right|_{\tl{q}=\tl{q}(p,q,\mu)}=-\textrm{arccosh}\left(\frac{A(\lm_1)+D(\lm_1)}{2}\right)-\textrm{arccosh}\left(\frac{A(\lm_2)+D(\lm_2)}{2}\right).
\ee
We remember that tr$(L)=A+D$ is the generating function of \emph{all} the conserved quantities of the system, see eq. (\ref{tra}); also the trace of $L$ is invariant under the map defined by the BTs, see eq. (\ref{intertwin}). As a result, the function $\Phi$ is invariant under the BTs, that is $\Phi(p,q,\mu)=\hat{\Phi}(\tl{p},\tl{q},\mu)$. Our maps then possess the spectrality property.

According to Theorem \ref{th1}, the BTs given by (\ref{D}) are the integral curves of the Hamiltonian system given by (\ref{newkmu}). 

As an explicit example let us take the simplest case, that is $N=2$. The trace of the Lax matrix is now given by
\be\label{tn2}
A(\lm)+D(\lm)=\lm^2+(p_1+p_2)\lm+p_1p_2 - 2\textrm{cosh}(q_1-q_2).
\ee 
In this case it is also quite straightforward to find explicitly a set of action-angle variables. A choice is given by the following relations:
\begin{equation}\label{th1th2} \begin{aligned}
&I_1=\frac{p_1+p_2}{2}, \qquad I_2=2\textrm{cosh}(q_1-q_2)+\left(\frac{p_1-p_2}{2}\right)^2,\\
&\theta_1=q_1+q_2, \qquad \theta_2= \frac{F(\frac{p_2-p_1}{2\sqrt{I_2-2}},k)}{\sqrt{I_2+2}}, 
\end{aligned}
\end{equation}
where $F(z,k)$ is the complete elliptic integral of the first kind, the modulus $k$ being defined by $k^2=\frac{I_2-2}{I_2+2}$. In these coordinates the expression (\ref{tn2}) becomes $A(\lm)+D(\lm)=(\lm+I_1)^2-I_2$.

According to Theorem \ref{th2}, taking into account expression (\ref{newkmu}), the recurrences defined by the BTs (\ref{rec}) discretize the continuous flow corresponding to the Hamiltonian
\begin{equation*}
\mathcal{H}\overset{N=2}{=}\int_{\lm_0+\mu}^{\lm_0-\mu}\textrm{arccosh}\left(\frac{(\lm+I_1)^2-I_2}{2}\right)d\lm ,
\end{equation*}
so that
\begin{equation}\label{AA2}
\tl{\theta}_j=\theta_j+\frac{\partial}{\partial I_j}\int_{\lm_0+\mu}^{\lm_0-\mu}\textrm{arccosh}\left(\frac{(\lm+I_1)^2-I_2}{2}\right)d\lm , \qquad j=1,2, 
\end{equation}
giving
\begin{equation}\label{citn} \begin{aligned}
&\tl{\theta}_1=\theta_1+\textrm{arccosh}\left(\frac{(\lm_0-\mu+I_1)^2-I_2}{2}\right)-\textrm{arccosh}\left(\frac{(\lm_0+\mu+I_1)^2-I_2}{2}\right),\\
&\tl{\theta}_2=\theta_2+\frac{1}{\sqrt{I_2+2}} \left(F(\frac{\lm_0-\mu+I_1}{\sqrt{I_2-2}},k)-F(\frac{\lm_0+\mu+I_1}{\sqrt{I_2-2}},k)\right).
\end{aligned}
\end{equation}

In these coordinates the BTs are linearized. Indeed, if we denote by $(I_j)_n$ and $(\theta_j)_n$ the $n^{\textrm{th}}$ iterates of the variables $I_j$ and $\theta_j$, as in (\ref{discr}), then from (\ref{citn}) we obtain
\begin{equation}\label{linBts} \begin{aligned}
& (I_1)_n=(I_1)_0, \\
& (I_2)_n=(I_2)_0, \\
&(\theta_1)_n=(\theta_1)_0+n\,\textrm{arccosh}\left(\frac{(\lm_0-\mu+I_1)^2-I_2}{2}\right)-n\,\textrm{arccosh}\left(\frac{(\lm_0+\mu+I_1)^2-I_2}{2}\right)\\
&(\theta_2)_n=(\theta_2)_0+\frac{n}{\sqrt{I_2+2}} \left(F\left(\frac{\lm_0-\mu+I_1}{\sqrt{I_2-2}},k\right)-F\left(\frac{\lm_0+\mu+I_1}{\sqrt{I_2-2}},k\right)\right)
\end{aligned}
\end{equation}

Also, comparing with the definition of $\theta_2$ in (\ref{th1th2}), it becomes clear that the BTs correspond to addition formulae for elliptic integrals. Indeed we can write
\begin{equation}\label{addform}
F(\tl{P},k)=F(P,k)+F(\lambda_{-},k)-F(\lambda_{+},k)
\end{equation}
where $\tl{P}=\frac{\tl{p}_2-\tl{p}_1}{2\sqrt{I_2-2}}$, $P=\frac{p_2-p_1}{2\sqrt{I_2-2}}$ and $\lambda_{\pm}=\frac{\lambda_0\pm\mu+I_1}{\sqrt{I_2-2}}$. For $N>2$ the BTs corresponds to addition formulae for hyper-elliptic integrals.
For similar results on addition formulae for Weierstra\ss  $\,\wp$ function or Jacobi elliptic functions see \cite{KV} and \cite{RZK}. 

\subsection{Numerics}\label{Numerics}
Formula (\ref{newkmu}) together with Theorem \ref{th2} give the following result: the BTs for the Toda lattice discretise the equations of motion given by the Hamiltonian
\begin{equation}\label{Hdis}
\mathcal{H}=-\int_{\lm_0-\mu}^{\lm_0+\mu}\textrm{arccosh}\left(\frac{\textrm{tr}\left(L(\lm)\right)}{2}\right)d\lm ,
\end{equation} 
where we recall that the trace of $L(\lm)$, given in (\ref{tra}), is the generating function of all the integrals.
For any function $\mathfrak{F}$ on the phase space, the corresponding continuous equation of motion is
\begin{equation}\label{eqdisc}
\dot{\mathfrak{F}}=\{\mathcal{H},\mathfrak{F}\}=\sum_k c_k\{H_k,f\}, \; \textrm{where} \;\; c_k\doteq -\int_{\lm_0-\mu}^{\lm_0+\mu}\frac{\lm^k}{\sqrt{\textrm{tr}(L(\lm))^2-4}}d\lm .
\end{equation}

Let us take the special case $N=3$. The three integrals of motion are given by:
\begin{equation} \begin{aligned}
& H_1 = p_1+p_2+p_3, \\
& H_2 = p_1p_2+p_2p_2+p_3p_1-e^{q_1-q_3}-e^{q_2-q_1}-e^{q_3-q_2},\\
& H_3 = p_1p_2p_3-p_2e^{q_1-q_3}-p_3e^{q_2-q_1}-p_1e^{q_3-q_2}.
\end{aligned}
\end{equation}
\begin{figure}
\centering
\includegraphics[scale=0.7]{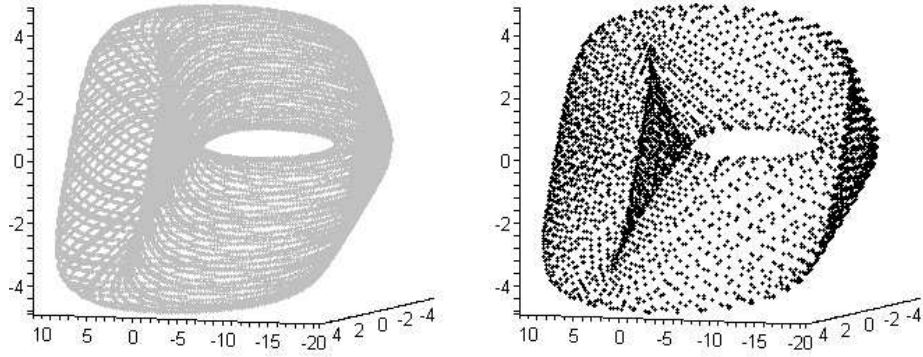}
\caption{\small{The initial conditions are $p_1=-20$, $p_2=10$, $p_3=1$,
$q_1=1$, $q_2=0$, $q_3=0.9$. The corresponding values for the constants $c_k$ are $c_0=-0.005$, $c_1=-0.048$ and $c_2=-0.512$.}}
\label{fig:1}
\end{figure}
\begin{figure}
\centering
\includegraphics[scale=0.7]{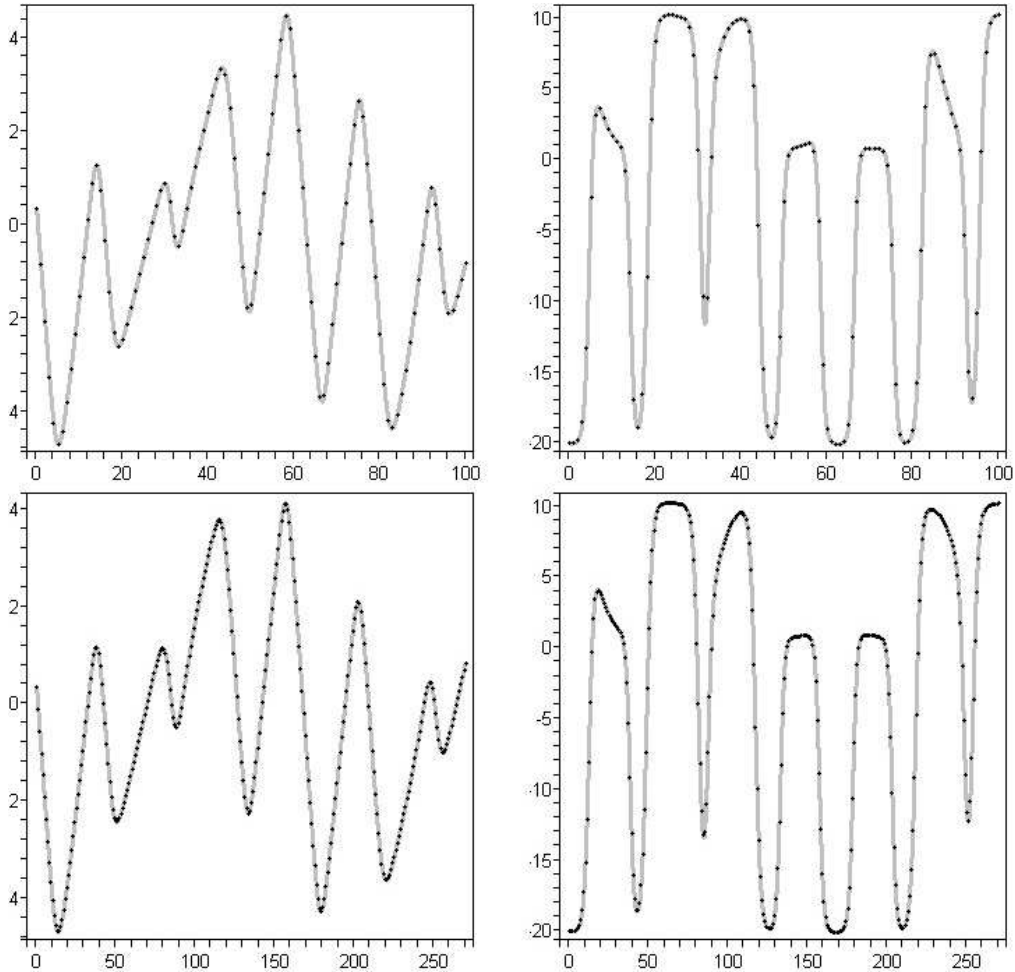}
\caption{\small{The values of $\mu$ are 5 (top) and 2 (bottom). The corresponding values for the constants $(c_0, c_1, c_2)$ are $(-0.005,-0.048,-0.512$ and $(-0.0018, -0.0183, -0.1849)$}}
\label{fig:2}
\end{figure}

Given the formulae (\ref{eqdisc}), it is possible to plot the continuous flow corresponding to some choice of the constants $c_k$ and the corresponding discrete flow given by the recurrences defined by the explicit BTs (from (\ref{new}), (\ref{xy}) and (\ref{lam0mu})). In order to make a comparison one has to choose the same initial conditions for the two flows. Once they are fixed (so that the conserved quantities $H_1$, $H_2$ and $H_3$ are given), it is possible to calculate, numerically or explicitly, the values of the constants $c_k$ knowing those of $\lm_0$ and $\mu$. Conversely, if there were three free parameters instead of two in the BTs, it would be possible to calculate their values knowing those of the constants $c_k$. Let us choose the values $\lm_0=10$ and $\mu=5$. A plot of the continuous flow given by (\ref{Hdiscr}) is given in Figure \ref{fig:1} on the left. On the right there are the first 3500 iterations of the BTs corresponding to the same initial conditions (both initial conditions and the values of the $c_k$ are given in the caption). The variables on the axes are $(p_3, q_2-Q, q_3-Q)$, where the center of mass $Q=\frac{q_1+q_2+q_3}{3}$, corresponding to a linear motion, has been subtracted to reveal the structure of a torus, in accordance with the Liouville-Arnold Theorem \cite{FM}.

In Figure \ref{fig:2} we report an example of the continuous trajectories (in grey) and the discrete ones (black dots): the top plots correspond to the variables $q_1-Q$ (on the right) and $p_1$ (on the left): the values of the initial conditions and of the parameters $\lambda_0$ and $\mu$ used are the same as for Figure \ref{fig:1}. For the two plots on the bottom we changed only the value of $\mu$ from 5 to 2: as explained in Section \ref{sec1} and as can be seen from the figures, a smaller value of the parameter $\mu$ corresponds to a smaller value of the time step size.

\appendix

\section{The two-parameter Darboux matrix}\label{Appendix}
First of all let us invert the transformations (\ref{ptp}), writing $\tl{p}_i$ and $\tl{q}_i$ in terms of $p$ and $q$. Recall that the Darboux matrix $D(\lm)$ is given by
\begin{equation}\label{d1} 
D(\lm) =  \left( \begin{array}{cc} \lm -\zeta -\alpha_1 \beta_1 & -\alpha_1 \\ \beta_1 & 1\end{array} 
\right). 
\end{equation}    

From (\ref{alphabeta}) we know that $\alpha_1=e^{q_1}$ and $\beta_1=e^{-\tl{q}_N}$. In order to write the transformations in explicit form we need to express $\beta_1$ as a function of the variables $p$ and $q$. This can be done noticing that the determinant of $D(\lm)$ calculated in $\lm=\zeta$ is zero. Indeed since $\det{\left(D(\zeta)\right)}=0$, $D(\zeta)$ has a one-dimensional kernel, spanned by $|\Omega>=(1,-\beta_1)^{T}$. The combination of the uniqueness of $|\Omega>$ (up to an overall factor) with (\ref{intertwin}) gives that $|\Omega>$ is also an eigenvector of $L(\zeta)$. Calling $v(\zeta)$ the corresponding eigenvalue, we have:
\begin{equation}\label{v1} \begin{aligned}
&A(\zeta)-\beta_1 B(\zeta)=v(\zeta),\\
&C(\zeta)-\beta_1 D(\zeta)=-v(\zeta)\beta_1 .
\end{aligned}
\end{equation}

After the elimination of $v(\zeta)$ we are left with $\beta_1^2 B(\zeta)-\beta_1(A(\zeta)-D(\zeta))-C(\zeta)=0$ defining $\beta_1$ in terms of only the dynamical variables without the tilde. At this point the relation (\ref{intertwin}) gives \emph{explicit} transformations.

Let us now consider the inverse transformations of (\ref{ptp}). They are defined by the inverse Darboux matrix $D^{-1}$ but since the relation (\ref{intertwin}) is homogeneous, we can take the adjugate matrix of $D$, say $\hat{D}$. So we have
\be\label{LhDhDL}
L(\lm,p,q)\hat{D}(\lm)=\hat{D}(\lm)L(\lm,\tl{p},\tl{q}),
\ee
with
\begin{equation}\label{hd} 
\hat{D}(\lm) =  \left( \begin{array}{cc} 1 & \alpha_1 \\ -\beta_1 & \lm -\zeta -\alpha_1 \beta_1\end{array} 
\right). 
\end{equation}

From (\ref{alphabeta}) we know that $\alpha_1=e^{q_1}$ and $\beta_1=e^{-\tl{q}_N}$ but in order to have explicit transformations we need to write $\alpha_1$ in terms of tilded variables. Again $\hat{D}(\zeta)$ has a kernel, spanned by $|\hat{\Omega}>=(\alpha_1,-1)^{T}$, that is also an eigenvector of $\tl{L}(\zeta)\doteq L(\zeta,\tl{p},\tl{q})$. So we have
\begin{equation}\label{v11} \begin{aligned}
&\tl{A}(\zeta)\alpha_1 -\tl{B}(\zeta)=w(\zeta)\alpha_1 ,\\
&\tl{C}(\zeta)\alpha_1-\tl{D}(\zeta)=-w(\zeta).
\end{aligned}
\end{equation}  

After the elimination of $w(\zeta)$ we are left with $(\alpha_1)^2 \tl{C}(\zeta)+\alpha_1(\tl{A}(\zeta)-\tl{D}(\zeta))-\tl{B}(\zeta)=0$. Note also that since $\det{\tl{L}}=\det{L}=1$ there are two possibilities for the eigenvalue $w(\zeta)$: either $w(\zeta)=v(\zeta)$ or $w(\zeta)=v(\zeta)^{-1}$, where $v(\zeta)$ is defined by (\ref{v1}). In order to have the inverse transformation of the previous one the correct choice is $w(\zeta)=v(\zeta)^{-1}$ as we will now show. This will fix the relative sign of the square roots appearing in the solutions of the quadratic equations defining $\beta_1$ and $\alpha_1$. 

The $(1,1)$ element of (\ref{LhDhDL}) gives $A(\zeta)-\beta_1 B(\zeta)=\tl{A}(\zeta)+\alpha_1 \tl{C}(\zeta)$. But from (\ref{v1}) we have $A(\zeta)-\beta_1 B(\zeta)=v(\zeta)$. Now it is simple to see that the equality $w(\zeta)v(\zeta)=1$, that can be written as
$
(\tl{D}(\zeta)-\alpha_1 \tl{C}(\zeta))(\tl{A}(\zeta)+\alpha_1 \tl{C}(\zeta))=1
$, 
is just the equation defining $\alpha_1$, because $\tl{A}\tl{D}=\tl{B}\tl{C}+1$.

We are ready to compose the two maps. We can define a new Darboux matrix $\Da(\lm, \lm_1,\lm_2)$ given by the product $\hat{D}(\lm,\lm_2)D(\lm,\lm_1)$, that is
\be\label{tiltil}\begin{split}
&\tl{L}D(\lm,\lm_1)=D(\lm,\lm_1)L, \qquad \tl{\tl{L}}\hat{D}(\lm,\lm_2)=\hat{D}(\lm,\lm_2)\tl{L}, \\
&\Longrightarrow \tl{\tl{L}}\hat{D}(\lm,\lm_2)D(\lm,\lm_1)=\hat{D}(\lm,\lm_2)D(\lm,\lm_1)L,
\end{split}\ee
with
\begin{equation}\label{D1D2} 
\hat{D}(\lm,\lm_2) =  \left( \begin{array}{cc} 1 & x_2 \\ -y_2 & \lm -\lm_2 -x_2 y_2\end{array} 
\right), \qquad D(\lm,\lm_1) =  \left( \begin{array}{cc} \lm -\lm_1 -x_1 y_1 & -x_1 \\ y_1 & 1\end{array} 
\right).
\end{equation}

From the relations (\ref{tiltil}) one has $y_1=y_2=e^{-\tl{q}_N}$. So we set $y_1=y_2\doteq y$. Also the product $\hat{D}(\lm,\lm_2)D(\lm,\lm_1)$ depends only on the difference $x_1-x_2$, so we set $x_1-x_2\doteq x$. The result is:
\begin{equation}\label{Dapp} 
\Da(\lm,\lm_1,\lm_2) =  \left( \begin{array}{cc} \lm-\lm_1-xy & -x \\ y(\lm_1-\lm_2+xy) & \lm -\lm_2 +x y\end{array} 
\right). 
\end{equation}

\section{The function $\Phi$}\label{AppendixB}
In this Appendix we obtain formula (\ref{newkmu}). 
From (\ref{vqqtmu}) we know that
$
\frac{\partial F}{\partial \mu}=\sum_k \ln\left(\frac{\eta_k+\mu}{\eta_k-\mu}\right).
$
In order to write this expression 
explicitly in terms of the set $(p_i,q_i)_{i=1}^{N}$ we note that the matrices $\da_j$ (\ref{daj}) are degenerate when $\lm =\lm_1$ and $\lm =\lm_2$. The kernels of $\da_j(\lm_1)$ are given by $|\omega_j(\lm_1)>=(x_j, \mu-\eta_j)^T$ and the relations (\ref{nlocal}) give 
$$
l_j(\lm_1)|\omega_j(\lm_1)>=\nu_{j+1}^1|\omega_{j+1}(\lm_1)>
$$
for some function $\nu_{j+1}^{1}$. More explicitly:
\be\label{rte}
(p_j+\lm_1)x_j-e^{q_j}(\mu-\eta_j)=\nu_{j+1}^1 x_{j+1}.
\ee
But from the combination of the $(2,1)$ and $(2,2)$ elements of (\ref{nlocal}) we have
\be\label{2122}
(p_j+\lm)x_j=e^{q_j}(\lm-\lm_0-\eta_j)+e^{\tl{q}_j}(\lm-\lm_0+\eta_{j+1}).
\ee
Evaluating the above equation at $\lm=\lm_1$ and comparing with (\ref{rte}) we can identify the function $\nu_{j+1}^{1}$ as
$$
\nu_{j+1}^1=\frac{e^{\tl{q}_j}(\mu+\eta_{j+1})}{x_{j+1}}
$$

Repeating the same argument for the kernels $|\omega_j(\lm_2)>=(-x_j, \mu+\eta_j)^T$ of $\da_j(\lm_2)$, we can write 
$$
l_j(\lm_2)|\omega_j(\lm_2)>=\nu_{j+1}^2|\omega_{j+1}(\lm_2)>
$$
for some function $\nu_{j+1}^2$, and again with the help of (\ref{2122}) evaluated at $\lm=\lm_2$ we find
$$
\nu_{j+1}^2=\frac{e^{\tl{q}_j}(\eta_{j+1}-\mu)}{x_{j+1}}.
$$

Now, due to the factorization of $L(\lm)$ into local matrices, the products $\prod_{k}\nu_{k}^1$ and $\prod_{k}\nu_k^2$ are eigenvalues of $L(\lm_1)$ and $L(\lm_2)$ respectively. But from the analysis of the eigenvalues of $L(\lm_1)$ and $L(\lm_2)$ we know that, when expressed in terms of the variables without tilde, we have (see (\ref{l1l2}) and (\ref{v1v2}))
\be
L(\lm_1)|\Omega_1>=v(\lm_1)|\Omega_1>, \qquad L(\lm_2)|\Omega_2>=v^{-1}(\lm_2)|\Omega_2>.
\ee
Collecting all of this together we see that $\frac{\partial F}{\partial \mu}$ is given by
\begin{equation}\label{spec}\begin{split}
&\left. \Phi(p,q,\mu)=\frac{\partial F}{\partial \mu}\right|_{\tl{q}=\tl{q}(p,q,\mu)}=\left.\ln\left(\frac{\prod_k \nu_k^1}{\prod_k \nu_k^2}\right)\right|_{\tl{q}=\tl{q}(p,q,\mu)}=\ln\left(v(\lm_1)\right)+\ln\left(v(\lm_2)\right)=\\
&\ln\left(\frac{A(\lm_1)+D(\lm_1)-\gm(\lm_1)}{2}\right)+\ln\left(\frac{A(\lm_2)+D(\lm_2)-\gm(\lm_2)}{2}\right).
\end{split}\end{equation}

The previous result can be also rewritten in terms of the inverse hyperbolic function arccosh($\cdot$), as in (\ref{newkmu}), to emphasize the dependence on the trace of the Lax matrix.

\vspace*{6mm}
\noindent\textbf{Acknowledgments}

\noindent
I wish to thank Andrew Hone for useful discussions, suggestions and remarks. Also, I would like to thank the referees for their helpful comments on the presentation of the paper. Finally, I wish to acknowledge the financial support of the \d'Istituto Nazionale di Alta Matematica'' (Italian National Institute for High Mathematics) as an INdAM-COFUND Marie Curie Fellow.

\end{document}